\documentstyle []{article}

\newcommand{\x}{\mbox{\boldmath $\xi$}}
\newcommand{\vi}{\hat{\mbox{\boldmath $\i$}}}
\newcommand{\vj}{\hat{\mbox{\boldmath $\j$}}}
\newcommand{\vk}{\hat{\mbox{\boldmath $\rm k$}}}
\newcommand{\vphi}{\hat{\mbox{\boldmath $\varphi$}}}
\newcommand{\nab}{\mbox{\boldmath $\nabla$}}

\begin{document}
 
\title{The tidally induced warping, precession and truncation of
accretion discs in binary systems: three--dimensional simulations}
\author{J.D. Larwood, R.P. Nelson, J.C.B. Papaloizou and 
C. Terquem}
 
\maketitle
 
\centerline{\bf Abstract}

\noindent We present the results of non linear, hydrodynamic simulations, in
three dimensions, of the tidal perturbation of accretion discs in
binary systems where the orbit is circular and not necessarily
coplanar with the disc mid--plane. The accretion discs are assumed to
be geometrically thin, and of low mass relative to the stellar mass so
that they are governed by thermal pressure and viscosity, but not
self--gravity. The parameters that we consider in our models are the
ratio of the orbital distance to the disc radius, $D/R$, the binary
mass ratio, $M_s/M_p$, the initial inclination angle between the orbit
and disc planes, $\delta$, and the Mach number in the outer parts of the
unperturbed disc,
${\cal M}.$
Since we consider non self-gravitating discs, these
calculations are relevant to protostellar binaries with separations below a few
hundred astronomical units.

\noindent For binary mass ratios of around unity and $D/R$ in the range 3 to 4,
we find that the global evolution of the discs is  governed primarily by the
value of ${\cal M}$. For relatively low Mach numbers (i.e. ${\cal M} =$
10 to 20) we find that the discs develop a mildly warped structure, are
tidally truncated,
and undergo a near rigid body precession at a rate which is in close agreement
with analytical arguments. For higher Mach numbers (${\cal M} \approx 30$),
the evolution is towards a considerably more warped structure, but the
disc none the less maintains itself as a long--lived, coherent entity. A
further increase in Mach number to ${\cal M} =50$ leads to a dramatic
disruption of the disc due to differential precession, since the sound
speed is too low to allow efficient communication between the disc's
constituent parts. Additionally, it is found that the inclination
angle between the disc and orbital angular momentum vectors evolves on
a longer timescale which is probably the viscous evolution timescale
of the disc.

\noindent The calculations are relevant to a number of observed astrophysical
phenomena,
including the precession of jets associated with young stars, the high
spectral index of some T~Tauri stars, and the light curves of X--ray
binaries such as Hercules X-1 which suggest the presence of precessing
accretion discs.

\section{Introduction}

Protostellar discs appear to be common around young stars. Furthermore
recent studies show that almost all young stars associated with low
mass star forming regions are in multiple systems (Mathieu~1994 and
references therein). Typical orbital separations are around 30
astronomical units (Leinert et al. 1993) which is smaller than the
characteristic disc size observed in these systems (Edwards et
al. 1987). It is therefore expected that circumstellar discs will be
subject to strong tidal effects due to the influence of binary
companions.
 
The tidal effect of an orbiting body on a differentially rotating disc
has been well studied in the context of planetary rings (Goldreich~\&
Tremaine~1981), planetary formation, and generally interacting binary
stars (cf. Lin~\& Papaloizou 1993 and references therein). In these
studies, the disc and orbit are usually taken to be coplanar (see
Artymowicz~\& Lubow~1994). However, there are observational indications
 that  discs and stellar orbits may not always be coplanar. For
instance, according to Corporon, Lagrange~\& Beust~(1996), the
pre-main sequence Herbig star TY~CrA is a triple system in which
the angle between the orbital plane of the third component and that of
the central binary has been inferred from spectroscopic observations to be
$\sim 70$ degrees. In addition a dusty shell is observed around the
binary (Bibo, The~\& Dawanas~1992). Although it is located at quite a distance
from the central binary (up to about 1000~astronomical units), this shell is
believed to be the remnant of an accretion disc. This would indicate that discs
and orbits are not necessarily coplanar in pre-main sequence systems. 
 In
addition, reprocessing of radiation from the central star by a warped
non coplanar disc has been suggested in order to account for the high
spectral index of some T~Tauri stars (Terquem \& Bertout 1993, 1996).
A dynamical study of the tidal interactions of a non coplanar disc is
of interest not only in the above contexts, but also in
relation to the possible existence of precessing discs which may define
the axes for  observed jets which apparently precess (Bally~\& Devine ~1994).

Linear calculations of the evolution of warped viscous discs were
performed by Papaloizou~\& Pringle~(1983). The related problem of the
propagation of bending modes in inviscid discs was considered by
Papaloizou~\& Lin~(1995). Papaloizou~\& Terquem~(1995) investigated
the tidal problem for an inviscid disc using linear perturbation
theory. Their results suggested that the disc would precess
approximately as a rigid body if the sound crossing time was smaller
than the differential precession frequency. They also estimated that
the tidally induced wave angular momentum flux arising from bending
distortions would lead to a small effective viscosity which may in
some rather extreme cases be significant for protostellar discs.
 
In this paper we extend the previous work and consider the full
nonlinear response of a tidally interacting disc which is not coplanar
with the binary orbit. We study the problem using a Smoothed Particle
Hydrodynamics (SPH) code originally developed by Nelson~\&
Papaloizou~(1993, 1994). In particular we aim to establish to what
extent strongly warped but thin discs may survive in close binary
systems and how the limitation of the disc size through the effect of
tidal truncation operates when the disc and binary orbit are not
coplanar. We find that the tidal truncation effect is only marginally
affected by lack of coplanarity. Also our model discs were able to survive in a tidally truncated condition while warped and undergoing rigid body precession provided that the Mach number in the disc was not too large (i.e. if the disc was not too thin). An additional result is
that the inclination of the disc was found to evolve only on a long
timescale likely to be the viscous timescale, as was indicated by the
linear calculations of Papaloizou~\& Terquem~(1995).
 
The plan of the paper is as follows: In section 2 we give the basic
equations. In section 3 we describe the orbital configuration and in
section 4 we discuss the precession of the disc in general terms. In
section 5 we review the numerical method and in sections 6 and 7 we
discuss its application to the disc problems considered and estimate
the amount of shear viscosity present.  Numerical results for discs
with mid-planes inclined at 45 and 90 degrees to the orbital plane are
presented for ratio of orbital separation to disc size between 3 and 4
in section 8. Mass ratios of unity and 0.3 have been
considered. Finally, we compare our results with some recent observations of
relevant astrophysical phenomena in section 9.

\section{Basic equations}
 
The equations of continuity and of motion applicable to a gaseous
viscous disc may be written

\begin{equation}  
\frac{d\rho}{dt} + \rho \nab \cdot {\bf v} = 0 ,
\label{cont} 
\end{equation}

\begin{equation}
\frac{d{\bf v}}{dt} = - \frac{1}{\rho} \nab P - \nab \Psi +
{\bf{S}}_{visc} 
\label{dvdt} 
\end{equation}

\noindent where 

\begin{displaymath}
{d \over dt} \equiv {\partial \over \partial t} +
{\bf v} \cdot \nab
\end{displaymath}

\noindent denotes the convective derivative, $\rho$ is the density, 
${\bf v}$ the velocity and $P$ the pressure. The gravitational
potential is $\Psi,$ and ${\bf{S}}_{visc}$ is the viscous force per
unit mass.

\noindent For the work described here, we adopt the polytropic 
equation of state

\begin{displaymath}
P= K \rho^{\gamma}
\end{displaymath}

\noindent where

\begin{displaymath}
c_s^2 = K \gamma \rho^{\gamma - 1}
\end{displaymath}

\noindent gives the usual associated sound speed, $c_s$. Here we take 
$\gamma=5/3,$ and $K$ is the polytropic constant. This corresponds to
adopting a fluid that remains isentropic throughout even though
viscous dissipation may occur. This means that an efficient cooling
mechanism is assumed.

\section{Orbital Configuration}

We consider a binary system in which the primary has a mass M$_p$ and
the secondary has a mass M$_s.$ The binary orbit is circular with
separation $D.$ The orbital angular velocity is $\omega.$ We suppose
that a disc orbits about the primary such that at time $t=0$ it has a
well defined mid-plane. We adopt a non rotating Cartesian coordinate
system $(x,y,z)$ centred on the primary star and we denote the unit
vectors in each of the coordinate directions by $\vi$, $\vj$ and $\vk$
respectively. The $z$ axis is chosen to be normal to the initial disc
mid-plane. We shall also use the associated cylindrical polar
coordinates $(r,\varphi,z).$

We take the orbit of the secondary to be in a plane which has an initial
inclination angle $\delta$ with respect to the $(x,y)$ plane. For a
disc of negligible mass, the plane of the orbit is invariable and does
not precess. We denote the position vector of the secondary star by
$\bf{D}$ with $D=|{\bf D}|.$ Adopting an orientation of coordinates
and an origin of time such that the line of nodes coincides with, and
the secondary is on, the $x$ axis at $t=0,$ the vector ${\bf D}$ is
given as a function of time by

\begin{equation}
{\bf D} = D \cos \omega t \;  \vi +
D \sin \omega t \cos \delta  \; \vj +
D \sin \omega t \sin \delta \; \vk .
\label{x}
\end{equation}

The total gravitational potential $\Psi$ at a point with position
vector ${\bf r}$ is given by

\begin{displaymath}
\Psi = - \frac{GM_p}{\mid \bf{r} \mid} -
\frac{GM_s}{\mid \bf{r} - \bf{D} \mid} +
{GM_s{\bf r} \cdot {\bf D}\over D ^3}
\end{displaymath}

\noindent where $G$ is the gravitational constant. The first dominant 
term is due to the primary, while the remaining perturbing terms are
due to the secondary. The last indirect term accounts for the
acceleration of the origin of the coordinate system. We note that a
disc perturbed by a secondary on an inclined orbit becomes tilted,
precesses and so does not maintain a fixed plane. Our calculations
presented below are referred to the Cartesian system defined above
through the initial disc mid-plane. However, we shall also use a
system defined relative to the fixed orbital plane for which the `$x$
axis' is as in the previous system and the `$z$ axis' is normal to
the orbital plane. If the disc were a rigid body its angular momentum
vector would precess uniformly about this normal, as indicated below.

\section{Precession Frequency}

In order to aid discussion of our numerical results, we consider in
general terms the response of the disc to the component of the time averaged
perturbing potential, ${\overline \Psi_s,}$ arising from the
secondary, that is odd in $z.$ This is the contribution responsible for
inducing disc precession. We
recall that to lowest order in $1/D$ this is (see Papaloizou~\& Terquem~1995):

\begin{equation}
{\overline \Psi_s} = - \frac{3}{4} \frac{GM_s}{D^3} r z 
\sin 2 \delta \sin \varphi .
\label{psib}
\end{equation}

\noindent  We take the unperturbed disc  to be in a state of Keplerian
rotation with local angular velocity $\Omega.$ Assuming that the basic equations
can be linearized, the Lagrangian displacement vector $\x$ will satisfy an
equation of the form (see Lynden-Bell~\& Ostriker~1967)

\begin{equation}
{\bf C}(\x) = -\nab {\overline \Psi_s}. 
\label{RESP} 
\end{equation} 

\noindent Here ${\bf C}$ is a linear operator, which needs to be 
inverted to give the response. When a barotropic equation of state
applies, and the boundaries are free, ${\bf C}$ is self-adjoint with
weight $\rho.$ This means that for two general displacement vectors
$\x({\bf r})$ and $\mbox{\boldmath $\eta$}({\bf r})$ we have
 
\begin{equation}
\int_V \rho \mbox{\boldmath $\eta$}^* \cdot
{\bf {C}} \left( \x \right) d\tau =
\left[ \int_V \rho \x^* \cdot {\bf {C}} \left(
\mbox{\boldmath $\eta$} \right) d\tau \right]^*
\end{equation}
 
\noindent where $^*$ denotes complex conjugate and the integral is 
taken over the disc volume V.
 
A general time averaged perturbing potential may be Fourier analyzed
in the $\varphi$ direction. We here consider the component with
azimuthal mode number $m=1.$ Because of the spherical symmetry of the
unperturbed primary potential, the unperturbed system is invariant under applying
a rigid tilt to the disc. This corresponds to the existence of rigid tilt mode
solutions to (\ref{RESP}) when there is no forcing (${\overline
\Psi}_{s}=0)$. For a rotation about the $x$ axis, the rigid tilt mode
is of the form

\begin{equation} 
\x = \x_{T} =
\left( {\bf r} \mbox{\boldmath $\times$} 
\vphi \right) \sin \varphi - \vphi z \cos \varphi
\label{tilt}
\end{equation} 

\noindent where $\vphi$ is the unit vector along the $\varphi$ 
direction. For time averaged forcing potentials $\propto \sin\varphi,$
the existence of the solution~(\ref{tilt}) results in an integrability
condition for (\ref{RESP}). When ${\bf C}$ is self-adjoint, this is

\begin{equation} 
\int_V \x_T \cdot \nab {\overline \Psi_s} 
\rho d\tau \equiv \int_V \vi \cdot
\left ({\bf r} \mbox{\boldmath $\times$} \nab 
{\overline \Psi_s}\right) \rho d\tau = 0. 
\label{INTCON} 
\end {equation}

\noindent The above condition is equivalent to the requirement that the $x$
component of the external torque vanishes. This will clearly not be
satisfied in the problem we consider.

To deal with this one may suppose that the disc angular momentum
vector precesses about the orbital angular momentum vector with
angular velocity $\mbox{\boldmath $\omega_p$}$ (Papaloizou~\&
Terquem~1995). This in turn is equivalent to supposing our coordinate
system rotates with angular velocity $\mbox{\boldmath $\omega_p$}$
about the orbital rotation axis. Treating the  Coriolis
force by perturbation theory produces an additional term on the right hand side
of~(\ref{RESP}) equal to 
$-2 r\Omega \mbox{\boldmath $\omega_p$} \mbox{\boldmath
$\times$} {\hat{\mbox{\boldmath $\varphi$}}}$. Using the modified force in
formulating~(\ref{INTCON}) gives the integrability condition as

\begin{equation} 
\omega_p \sin \delta \int_V r^2 \Omega \rho d\tau =
\int_V \vi \cdot \left ({\bf r} \mbox{\boldmath $\times$} 
\nab {\overline \Psi_s} \right) \rho d\tau , 
\label{PRECESS} 
\end{equation}

\noindent with $\omega_p = |\mbox{\boldmath $\omega_p$}|.$ 
Equation~(\ref{PRECESS}) gives a precession frequency for the disc
that would apply if it were a rigid body. However, approximate rigid
body precession is only expected to occur if the disc is able to
communicate with itself, either through wave propagation or viscous
diffusion, on a timescale less than the inverse precession frequency
(see for example Papaloizou~\& Terquem~1995 and below). Otherwise, a
thin disc configuration may be destroyed by strong warping and
differential precession.

\section{Numerical Method}

The  set of basic equations~(\ref{cont}) and~(\ref{dvdt}) are solved
numerically using an SPH code
(Lucy~1977, Gingold~\& Monaghan~1977), developed by Nelson~\&
Papaloizou~(1994), which uses a conservative formulation of the method
that employs variable smoothing lengths. 
A suite of test calculations illustrating the
accurate energy conservation obtained with this method is described by
Nelson~\& Papaloizou~(1994), and additional tests and calculations are 
presented in Nelson~(1994).

\subsection{Hydrodynamical Equations}

\subsubsection{ Smoothed Functions} 

In SPH one works with smoothed quantities defined by the expression

\begin{equation} 
\langle f({\bf r}) \rangle = \int f({\bf r'}) 
W \left( \left| {\bf r} - {\bf r'} \right|,h \right) d^{3}{\bf r'} 
\label{sf1} 
\end{equation} 

\noindent where $f({\bf r})$ is an arbitrary function, $W$ is a 
smoothing kernel, and $h$ is the smoothing length. Because the
function $f({\bf r})$ is defined only at particle positions, a
Monte-Carlo representation of equation~(\ref{sf1}) is used in the form

\begin{equation} 
\langle f({\bf r}_{i}) \rangle =
\sum^{N}_{j=1} \frac{m_{j}}{\rho_{j}} f({\bf r}_{j}) 
W \left( \left| {\bf r}_{i} - {\bf r}_{j} \right|, h \right) 
\label{sf2} 
\end{equation} 

\noindent where $m_{j}$ is the mass of particle $j,$ $\rho_{j}$ is 
the density at the location of particle $j,$ and the sum is
over all $N$ particles. If the function
$f({\bf r})$ is chosen to be the density, $\rho({\bf r})$, then we obtain

\begin{equation} 
\rho({\bf r}_{i}) = \sum_{j=1}^{N} m_{j} 
W \left( \left| {\bf r}_{i} - {\bf r}_{j} \right|, h \right)
\label{den1} 
\end{equation}

\noindent where the angled brackets have been dropped for the sake of 
brevity, with the understanding that smoothed functions are used
throughout in SPH.

When variable smoothing lengths are employed each particle, $i,$ has
an associated smoothing length $h_{i}.$ In order to conserve momentum
the kernel is symmetrised with respect to particle pairs using the
procedure of Hernquist~\& Katz~(1989) which  leads to the replacement:

\begin{equation} 
W_{ij} \rightarrow \frac{1}{2} \left[ 
W \left( \left| {\bf r}_{i} - {\bf r}_{j} \right|, h_{i} \right) +
W \left( \left| {\bf r}_{i} - {\bf r}_{j} \right|, h_{j} \right) \right] 
\label{Wij} 
\end{equation}

The smoothing function used is the commonly employed spline kernel initially
suggested by Monaghan~\& Lattanzio~(1985).

\noindent Then the smoothed density~(\ref{den1}) becomes

\begin{equation} 
\rho({\bf r}_{i}) = \sum_{j=1}^{N} m_{j} \frac{1}{2} \left[ 
W \left( \left| {\bf r}_{i} - {\bf r}_{j} \right|, h_{i} \right) + 
W \left( \left| {\bf r}_{i} - {\bf r}_{j} \right|, h_{j} \right) \right].
\label{den2}  
\end{equation}
 
\subsubsection{Equations of Motion}

The equation of motion for particle $i$ is

\begin{equation} 
m_i \frac{d{\bf v}_i}{dt} = {\bf F}_{P,i} + {\bf F}_{ G,i}
+ {\bf{S}}_{visc,i} . 
\label{dvdti}
 \end{equation}

\noindent Here we have denoted the forces  on particle $i$ due to 
pressure, gravity and viscosity as ${\bf F}_{P,i}, {\bf F}_{G,i},$ and
${\bf{S}}_{visc,i}$ respectively.  

\subsubsection{The pressure force}

The pressure force is given by the expression
 
\begin{eqnarray}
{\bf F}_{P,i} & = & - \sum_{j=1}^{N} m_{i} m_{j}
\left( \frac{P_{i}}{\rho_{i}^{2}} + \frac{P_{j}}{\rho_{j}^{2}} \right) 
\frac{1}{2} \left[
\left. \frac{\partial W({\bf r}_{ij},h_i)}{\partial |{\bf r}_{ij}|} 
\right|_{h_i const}
\frac{{\bf r}_{ij}}{| {\bf r}_{ij} |}
+ \left. \frac{\partial W({\bf r}_{ij},h_j)}{\partial |{\bf r}_{ij}|}
\right|_{h_j const}
\frac{{\bf r}_{ij}}{| {\bf r}_{ij} |} \right]
\nonumber  \\
& &  - \sum_{k=1}^{N} \sum_{j=1}^{N} m_k m_j
\left( \frac{P_{k}}{\rho_{k}^{2}} \right) \frac{1}{2} \left[
\frac{\partial W({\bf r}_{kj},h_k)}{\partial h_k}  
\frac{\partial h_k}{\partial {\bf r}_{i}} + 
\frac{\partial W({\bf r}_{kj},h_j)}{\partial h_j}
\frac{\partial h_j}{\partial{\bf r}_{i}} \right] 
\label{gradp3}
\end{eqnarray}

\noindent where we adopt the notation ${\bf r}_{kj} \equiv 
{\bf r}_k - {\bf r}_j$. A full derivation of this form of the
pressure force equation is given in Nelson~\& Papaloizou~(1994).
The first part of the above pressure force,
that does not involve derivatives of the smoothing length, is of the
usual symmetrised standard form (Hernquist~\& Katz~1989). In order to
evaluate the second part, we need to specify the way the smoothing
length depends on the particle coordinates.

\noindent In order that the system conserve linear and angular momentum
as well as energy, we take the $h_k$ to be functions only of the
absolute distances between particles. Here, we shall adopt the
prescription given in Nelson~\& Papaloizou~(1994), namely we first
find the ${\cal N}_{TOL}$ nearest neighbours and then calculate $h_i$
from this list of nearest neighbours according to:

\begin{equation} 
h_i = \frac{1}{N_{far}} \sum_{n=1}^{N_{far}} \frac{1}{2} 
| {\bf r}_i - {\bf r}_n | 
\label{hn6} 
\end{equation}
 
\noindent where the summation is over particle $i$'s $N_{far}$ most 
distant nearest neighbours. For the calculations presented here, we
have used ${\cal N}_{TOL} =45 $ and $N_{far}=6.$ Then the value of
$2h_i$ is equal to the mean distance to particle $i$'s six most
distant nearest neighbours.
 
\subsubsection {The addition of viscosity}

In order to stabilize the calculations in the presence of shocks, we
add an artificial viscous pressure term, $\Pi_{ij}$ 
(Monaghan~\& Gingold~1983), in only the leading term of equation~(\ref{gradp3}),
such that

\begin{displaymath}
\frac{P_i}{\rho_{i}^{2}} + \frac{P_j}{\rho_{j}^{2}} \rightarrow
\frac{P_i}{\rho_{i}^{2}} + \frac{P_j}{\rho_{j}^{2}} + \Pi_{ij}.
\end{displaymath}

\noindent With this addition

\begin{displaymath}
{\bf F}_{P,i} \rightarrow {\bf F}_{P,i} + {\bf{S}}_{visc,i}.
\end{displaymath}
The artificial viscous pressure given by Gingold~\& Monaghan~(1983) is

\begin{equation} 
\Pi_{ij} = \frac{1}{{\bar\rho}_{ij}} \left( 
- \alpha \mu_{ij} {\bar c}_{ij} + \beta \mu^{2}_{ij} \right) 
\label{Pi}
\end{equation} 

\noindent where the notation ${\bar A}_{ij} = \frac{1}{2}(A_{i}+A_{j})$ 
has been used, ${\bar c}_{ij}$ is the mean sound speed evaluated at particles
$i$ and $j$, and

\begin{equation} 
\mu_{ij} = \left\{ \begin{array}{cl} \displaystyle
{\bar h_{ij}} \frac{{\bf v}_{ij}.{\bf r}_{ij}}{r_{ij}^{2} + \eta^{2}} &
\mbox{if ${\bf v}_{ij}.{\bf r}_{ij} < 0 $} 
\\ 0 & \mbox{otherwise.}
\label{muij} 
\end{array} \right. 
\end{equation} 

\noindent Here, the notation ${\bf v}_{ij}=({\bf v}_{i}-{\bf v}_{j})$ 
has been used, and $\eta^{2}=0.01{\bar h}^{2}_{ij}$ prevents the
denominator from vanishing. For all calculations reported here, we
adopted $\alpha = 0.5$ and $\beta =0$ (cf. Artymowicz~\& Lubow~1994).

\subsubsection{Time stepping}

The particle equations of motion were integrated using the standard 
leap frog method. The time step size was determined using the criterion
presented in Nelson~\& Papaloizou~(1994), with the value of the Courant 
number being set to ${\cal Q}=0.3$.

\section{Initial set up}

A disc containing 17500 identical particles was initially set in orbital
motion about the primary alone. This was assumed to act through a
softened gravitational potential such that

\begin{displaymath}
\Psi_{p} = - {GM_p \over \sqrt{r^2 +b^2}} ,
\end{displaymath}

\noindent where $b$ is the softening length. With this prescription 
the disc could be extended to $r=0$ as was the case for the
calculations presented here. We adopt units such that the primary mass
$M_p=1$, the gravitational constant $G=1$, and the outer disc radius
$R=1$. In these units the adopted softening parameter $b=0.2$ and the time
unit is $\Omega(R)^{-1}.$ The
particle positions were initialised by dividing the disc into 50
annuli of equal width and placing in each one the number of equal mass
particles required to maintain a constant $\Sigma$ distribution over
the whole disc, where $\Sigma$ is the surface density averaged over an
annulus. The particles were then separated into seven equally spaced
planes, giving a total uniform semi-thickness $H.$ 

\noindent We note that in
 standard axisymmetric  accretion disc theory, vertical hydrostatic 
equilibrium at a general radius
 gives 
 $c_s(r) \sim \Omega(r) H(r).$  In the outer parts of the disc, 
this is equivalent to
${\cal M}=(H/R)^{-1}$, where we define the Mach number ${\cal M}$ as:

\begin{displaymath}
{\cal M}\equiv v_\phi  /c_s ,
\end{displaymath}

\noindent where  the azimuthal velocity  $v_\phi,$ and $c_s$ are 
evaluated  in  the  outer regions  of the disc  just interior to 
the  disc edge. Hence we say that
the Mach number ${\cal M}$ parameterises the hydrostatic state of the
disc. The value of $H$ was estimated so as to give vertical
hydrostatic equilibrium in the neighbourhood of the outer disc
boundary  for the required
 Mach number. 

\noindent After determining $H,$ the particle positions in each plane section
for each annulus were generated pseudo-randomly by computer. The particle
velocities were set to be initially azimuthal and equal to $r \Omega,$
the circular velocity, such that, in our units

\begin{displaymath}
\Omega = {1 \over \left( r^2 + b^2\right)^{3/4}} .
\end{displaymath}

\noindent Finally the required mean Mach number in the outer regions of the
disc, ${\cal M}$, was made consistent with the equation of state by suitably
adjusting the polytropic constant $K$, since at any location ${\cal M}^2=
(r^2\Omega^2)/(K\gamma\rho^{\gamma -1 })$.

\noindent For all of the calculations described here the smoothing length
was less than the disc semi-thickness so that the latter was
determined by genuine pressure effects. Only for the largest value
considered (${\cal M}=50$) were the smoothing length and
semi-thickness comparable.

The disc models were relaxed, before introducing the secondary, for
approximately two orbital periods at the outer boundary. During this
period the outside edge of the disc expanded by about 20 percent to
1.2. This effect was mostly due to a pressure-driven expansion into
the surrounding vacuum, near the outer edge, and a small viscous
expansion. After relaxation, the Mach number was found to be approximately 
constant through the disc and  the expansion at the outside was found to
proceed through the action of viscosity alone.

\section{Viscosity Calibration}

Although in a formal sense the artificial viscosity specified
by~(\ref{Pi}) is a bulk viscosity designed to handle shocks, its
practical implementation induces a shear viscosity $\nu = C_v \alpha
c_s h,$ with $C_v$ being a constant usually in the range 0.1 to 1
(Artymowicz \& Lubow 1994). This
leads to angular momentum transport and the standard viscous evolution
of an accretion disc (Lynden-Bell~\& Pringle~1974).  In studies of the
type presented here where discs undergo angular momentum exchange
through tidal interaction with orbiting secondaries, it is important
to know the rate at which viscosity transports angular momentum so
this may be compared with other effects. In the calculations presented
here it is evident that in many cases, disc expansion arising from
outward transport of angular momentum is halted by tidal
truncation. This effect, well known in the coplanar case (Lin~\&
Papaloizou~1993), is seen here when the disc and binary angular
momenta are not aligned.

\noindent In order to estimate the magnitude of the shear viscosity
operating in our disc models we studied some freely expanding disc
models for up to fifty time units. For our variable smoothing length
prescription we expect that $h \propto \rho^{-1/3}.$ Then, as for our
equation of state $c_s \propto \rho^{1/3},$ we expect $\nu$ to be
constant for our disc models. To specify the magnitude of $\nu$ we
write $\nu =\alpha_{ss} c_s^2(R)/\Omega (R),$ where $\alpha_{ss}$
corresponds to the well known Shakura~\& Sunyaev~(1973) $\alpha$
parameterization. However, it is applicable only at the outside edge
of the disc.

The standard diffusion equation for the evolution of a viscous disc
with $\nu$ a constant is (Lynden-Bell~\& Pringle~1974)

\begin{equation}
{\partial \Sigma \over \partial t} =
{-\nu \over r}{\partial \over \partial r} \left\{
\left[ \left( r^2 \Omega \right)' \right]^{-1} 
{\partial \over \partial r}
\left( \Sigma r^3 \Omega' \right) \right\} 
\label{diffe}
\end{equation}
where a prime denotes differentiation of a function with respect to $r$.

\noindent We found that the evolution of a freely expanding SPH disc, 
with ${\cal M}=10,$ followed~(\ref{diffe}) to high accuracy for $0 < r
< 1$ and $0 < t < 50.$ This was found to be the case immediately after
set up with $\Omega$ evaluated using the softened primary potential,
when $\alpha_{ss}=0.03$ near the outer boundary of the disc. This
corresponds to a global viscous evolution timescale of $4000$
time units. Although this is a longer time than any of our models were
evolved for, the time required to establish tidal truncation after a
radial expansion of $20$ percent is significantly shorter, being about
160 units for ${\cal M}=20$.

We also comment that a value of $\alpha_{ss}=0.03$ is about two orders
of magnitude larger than that expected to be associated with tidally
induced inwardly propagating waves (Spruit~1987, Papaloizou~\&
Terquem~1995). Accordingly, it is expected that tidal truncation will
instead occur through strong nonlinear dissipation near the disc's
outer edge (Savonije, Papaloizou~\& Lin~1994). The large viscosity
of the disc models considered here will damp inwardly propagating
waves before they can propagate very far.

\section{Numerical Results}

We have considered the evolution of the disc models set up according
to the procedure outlined above. Details of the models are given in
Table 1. In the absence of the secondary star the models were
characterized only by the mean Mach number in the outer regions of the
disc, ${\cal M}$. After the relaxation period of about two rotation
periods at the outer edge of the disc, the time was reset to zero and
the secondary was introduced in an inclined orbit, moving in a direct
sense, crossing the $x$ axis at $t=0.$ The parameters of the binary
orbit are given in Table~\ref{table1}. In some cases the full
secondary mass was included immediately corresponding to a sudden
start. However, for strong initial tidal interactions such as those
that occur when $D/R=3, M_s/M_p=1$, this can result in
disruption of the outer edge of the disc with a small number of particles
being ejected from the disc. It was found that this
could be avoided by using a
`slow start' in which $M_s$ was built up according to

\begin{displaymath}
M_s(t) = M_s \left( 1 - A \exp(-at) \right) ,
\end{displaymath}
 
\noindent where we used $A=\exp(-0.004)$, and $a=0.4$. Consequently,
the total companion mass $M_s$ was present in the calculations after a time
of $\approx 10$.

Models were continued for the run times indicated in
Table~\ref{table1}. Most binary systems had $D/R=3,$ but cases with
$D/R=3.6$ and $4$ were also considered. The general finding was that a
disc with an initial angular momentum vector inclined to that of the
binary system tended to precess approximately as a rigid body, with a
noticeable but small warp if ${\cal M}$ was not too large. In such
cases only small changes in the inclination angle between the angular
momentum vectors were found over the run time. This is consistent with
the expectation from Papaloizou~\& Terquem~(1995) that the timescale
for evolution of the inclination in such cases should be comparable to
the viscous evolution timescale of the disc, assuming outward disc
expansion is prevented by tidal interaction.

\begin{table}[h]
\caption{Model discs}
\begin{center}
\begin{tabular} {cccccccc} \hline \hline
Model & $M_s/M_p$ & $D/R$ & $\delta$ 
& ${\cal M}$ & Start & Run time & \\ 
\hline
       &     &     &         &     &        &       & \\
$\; 1$ & 1   & 3   & $\pi$/4 & 20  & slow   & 310.0 & \\
       &     &     &         &              &       & \\
$\; 2$ & 1   & 3   & $\pi$/4 & 25  & slow   & 217.7 & \\
       &     &     &         &     &        &       & \\
$\; 3$ & 1   & 4   & $\pi$/4 & 20  & sudden & 330.0 & \\
       &     &     &         &     &        &       & \\
$\; 4$ & 1   & 3.6 & $\pi$/4 & 10  & sudden & 152.0 & \\
       &     &     &         &     &        &       & \\
$\; 5$ & 1   & 3.6 & $\pi$/2 & 10  & sudden & 156.0 & \\
       &     &     &         &     &        &       & \\
$\; 6$ & 0.3 & 3.6 & $\pi$/4 & 10  & sudden & 84.0  & \\
       &     &     &         &     &        &       & \\
$\; 7$ & 1   & 3   & $\pi$/2 & 20  & slow   & 328.5 & \\
       &     &     &         &     &        &       & \\
$\; 8$ & 1   & 4   & $\pi$/2 & 20  & slow   & 228.0 & \\
       &     &     &         &     &        &       & \\
$\; 9$ & 1   & 3   & $\pi$/4 & 30  & slow   & 397.9 & \\
       &     &     &         &     &        &       & \\
$\;10$ & 1   & 3   & $\pi$/4 & 50  & slow   & 50.0  & \\
       &     &     &         &     &        &       & \\
$\;11$ & 1   & 3.6 & 0.0     & 10  & sudden & 150.8 & \\
       &     &     &         &     &        &       & \\
\hline \hline
\end{tabular}
\end{center}
\label{table1}
\end{table}

The calculations presented here use a coordinate system which is based
on the initial disc mid-plane. However, as the disc precesses, the
mid-plane changes location with time. It is then more convenient to
use a coordinate system $(x,y_o,z_o)$ based on the fixed orbital
plane, the $z_o$ axis coinciding with the orbital rotation axis.
We shall refer to these
as `orbital plane coordinates'. We locate the inclination angle
$\iota$ (equal to $\delta$ at $t=0$) between the disc and binary orbit
angular momentum vectors through

\begin{displaymath}
\cos \iota = { {\bf J}_D\cdot  {\bf J}_O \over | {\bf J}_D||{\bf J}_O|} .
\end{displaymath}

\noindent Here, ${\bf J}_O$ is the orbital angular momentum. 
The disc angular momentum is ${\bf J}_D=\sum_j {\bf J}_j,$ where the
sum is over all disc particle angular momenta ${\bf J}_j.$
 
\noindent A precession angle $\beta_p,$ measured in the orbital plane 
can be defined through

\begin{displaymath}
\cos \beta_p = - { ({\bf J}_D \mbox{\boldmath $\times$} 
{\bf J}_O)\cdot {\bf u} \over | {\bf
J}_D{\mbox{\boldmath $\times$} \bf J}_O| |{\bf u}|} 
\end{displaymath}

\noindent where ${\bf u}$ may be taken to be any fixed reference 
vector in the orbital plane. We take this to point along the $y_o$
axis such that initially $\beta_p = \pi /2$ in all cases. For retrograde
precession of ${\bf J}_D$ about ${\bf J}_O$ the angle $\beta_p$ should
initially decrease as is found in practice. The period of revolution
for rigid body precession of a disc is $2\pi /\omega_{p}$ with, from
equation~(\ref{PRECESS}):

\begin{equation}
\omega_p = - \left( {3GM_s \over 4D^3} \right) \cos \delta
{\int^R_0 \Sigma r^3 dr \over \int^R_0 \Sigma r^3 \Omega dr} .
\end{equation}

\noindent For a disc with constant $\Sigma$ and radius $R,$ this gives

\begin{equation}
{\omega_p \over \Omega(R)}
= -{15 M_s R^3 \over 32 M_p D^3} \cos \delta .
\label{Prc}
\end{equation}

\noindent However, near rigid body precession could only be expected 
if there is good communication, either through wave propagation or
viscous processes, throughout the disc. As long as $\alpha_{ss} <
{\cal M}^{-1},$ which is at least marginally satisfied in all models
except model 10,
sound waves may propagate (Papaloizou~\& Pringle~1983). The ability
of sound to propagate throughout the disc during a precession time
implies approximately that

\begin{equation} 
\frac{H}{R} > \frac{|\omega_p|}{\Omega(R)} . 
\label{cond} 
\end{equation}

\noindent When $M_s/M_p=1$, $D/R=3$ and $\delta =\pi /4$, 
equation~(\ref{Prc}) gives $\omega_p/\Omega(R) = 0.012$ corresponding
to a precession period of 512 in our time units. Our results were
consistent with the condition~(\ref{cond}) to within a factor of two
in that models with ${\cal M} < 25, \delta = \pi /4$ and $D/R > 3$
showed modest warps and approximate rigid body precession. Model~9
with ${\cal M} = 30$ showed severe warping and a more complex
precessional behaviour while model~10 with ${\cal M}=50 $ appeared to
be disrupted by differential precession.

We now describe in more detail the results from some of our runs. We
present particle projection plots for each of the Cartesian planes
using obital plane coordinates. In all such figures a fourth
`sectional plot' is also included in which only particles such that
$-0.05 < y_{o}< 0.05$ are plotted.

\subsection{Models 1 \& 3}
 
A projection plot for model~1 is shown in
Fig.~(\ref{fig1}) 
near $t=0$ when the disc is
unperturbed. Note that the disc appears as edge on and inclined at $45$
degrees in the $(y_o ,z_o)$ plane.
The time dependence of the angles $\iota$ and $\beta_p$ is plotted in
Fig.~(\ref{fig2}). It may be seen that there is little change in $\iota$ during
the whole run. On the other hand $\beta_p$ decreases approximately linearly,
corresponding to uniform precession (note that $\beta_p$ is plotted as positive
rather than negative for the latter section of this plot). The inferred
precession period is around 600 units, in reasonable agreement with the value of
512 units obtained from equation~(\ref{Prc}). Fig.~(\ref{fig3}) shows a
projection plot at time 156.6 after the disc has precessed through about 90
degrees in a retrograde sense. This amount of precession is evident because at
this time the projection in the $(x,z_o)$ plane shows  the disc to be almost
edge on, and inclined at about $45$ degrees. Note that the disc remains thin. 
Fig.~(\ref{fig4}) shows another projection at $t=228.8$. This is taken at a
viewing angle such that the disc is almost edge on. Particles from the disc at
time $t=0$ are also projected. It is apparent that at this time the disc plane
is approximately orthogonal to the original one. The later disc is also clearly
of smaller size demonstrating the effect of tidal truncation. We note that
the same configuration, viewed from an angle at which the original disc
appears edge on, 
indicates that the apparent smaller size of the evolved disc is {\em not} the
result of projecting the short  axis of an elliptical disc. The mean radius of
the evolved disc is indeed smaller
than the original disc radius due to tidal truncation.

\noindent Fig.~(\ref{fig5}) shows a  plot similar to Fig.~(\ref{fig3})
at $t=297.8$, near the end
of the run when the disc has precessed through about 180 degrees. This
amount of precession is demonstrated by the fact that the disc appears
almost edge on in the $(y_o ,z_o)$ plane just as it did at time $t=0.$ However,
its plane is inclined at about $90$ degrees to the original disc plane. At this stage
our results indicate that the disc has attained a quasi-steady configuration
as viewed in a frame that precesses uniformly about the orbital rotation axis.
In this run, as in others, the disc develops a warped structure that initially
grows in magnitude but then levels off.  The sectional plot
in Fig.~(\ref{fig5}) indicates that
the disc has developed a modest warp in this case. 

\noindent For comparison with model~1, Fig.~(\ref{fig6}) shows a 
corresponding projection for
model~3 with $D/R=4$ at $t=330.0$. As expected the precession rate is
about half as fast in this case, but the sectional plot indicates that
the warp is not less pronounced than in model~1. 
This may be related to the fact that because of the more distant companion,
the disc is able to expand to a larger size. We also note that the
linear calculations of  Papaloizou~\& Terquem~(1995) indicated that the low
frequency components of the perturbation do not diminish in their effectiveness
until $D/R \sim 8$. 

\subsection{Models~9,~2,~\& 10}

In contrast to models~1 and~4, in which the Mach numbers are
${\cal M} = 20$ and ${\cal M} =10$ respectively,
the behaviour of model~9 with ${\cal
M}=30$ but similar orbital parameters
is considerably more complex. This disc develops a strong 
warp such that the
inner and outer parts of the disc try to separate. A projection plot
is shown at $t=397.9$ in Fig.~(\ref{fig7}). The inner part of the disc
seems to occupy a different plane from the outer part. The outer part
was found to precess like a rigid body at the expected rate, and it
tended to drag the inner section behind it. This is indicated in
Fig.~(\ref{fig8}) where we plot the angle $\beta_p$ calculated using
the outer disc section with $r> 0.5$ only and also the same angle
calculated using  only the inner section with $r< 0.3.$ These do not
correspond exactly but the angle associated with the inner segment
progresses at a variable rate indicating coupling to the outer section. The
angle $\iota$ associated with the two sections is also plotted. This
becomes significantly smaller for the inner segment consistent with
the existence of a large amplitude warp. We note that the relatively
larger inclination associated with the outer segment enables a closer
matching of the precession frequencies associated with the two
segments and aids coupling, due to the presence of large 
pressure gradients induced by the strong warping. 
Each segment of this model remained thin
throughout the run. 

\noindent For comparison with model~9, Fig.~(\ref{fig8bis}) shows the
evolution of $\beta_p$ and $\iota$ for model~2, which is the same as
model~9 but with ${\cal M}=25.$ This is an intermediate case between
model~9 and model~1. The precession is slightly differential at the
beginning of the run, but the inner part of the disc remains coupled to the
outer part, and the precession becomes uniform after about 150 time
units. Since the inner part adjusts its precession frequency to the
outer part, it changes slightly its inclination.  

\noindent In the case of model~10, which had a 
Mach number of ${\cal M}=50$, severe
disruption of the disc due to differential precession was found to occur.
The thickness of the disc was such that efficient communication between
different parts of the disc was unable to occur, and consequently a
well defined, coherent warped structure could not be maintained. Instead,
a decidedly non disc--like morphology was obtained by the time the 
calculation had been evolved for $t=50$ units.


\subsection{Models~4~\& 11}

We also studied relatively thick disc models with ${\cal
M}=10.$ For example model~4 had $D/R=3.6,$ and $\delta=\pi/4.$ These
models also precessed like rigid bodies with barely discernable
warps. We plot the behaviour of $\iota$ and $\beta_p$ (calculated for
the whole disc) in Fig.~(\ref{fig10}). In this case the precession
period is estimated to be 873 units, again consistent
with the value of 884 units predicted 
by equation~(\ref{Prc}). We note that the agreement between 
theory and the numerical
calculations is better in the ${\cal M}=10$ than for the higher Mach number 
cases calculation because the
larger sound speed leads to improved communication throughout the disc, allowing
it to precess more as a rigid body.
Here $\iota$ shows little evolution over the
duration of the run. A projection plot is shown at $t=105.1$ in
Fig.~(\ref{fig9}). This figure illustrates truncation of the disc at $R=1$ as a
freely expanding disc was found to expand significantly further. For
comparison we ran the same model but with $\delta=0$ (coplanar). A
projection plot for this model~11 at $t= 147.3$ is shown in
Fig.~(\ref{fig11}). 
It was found that the discs in models~4 and~11 are
of about the same size and thickness. Thus the tidal truncation radius
for $\delta = \pi /4$ is similar but slightly larger than in the coplanar case. 

\subsection{Model~6}

We
also ran model~6 which had the same parameters as model~4, but 
with a mass ratio of $M_s/M_p =0.3$.
The lower secondary mass results in a bigger disc, and
a precession rate about three times slower, as one would predict from
equation~(\ref{Prc}). A projection plot at $t
=84 $ is given in Fig.~(\ref{fig12}), in which there is little sign of a warp.

\subsection{Models 5, 7 \& 8}

In addition to the cases described above we also ran models with
$\delta = \pi /2.$ In this case the disc plane remains in a stable
position apart from some oscillatory bending and rotation. This is
again consistent with inclination evolution occuring on the global viscous
timescale. These discs also remained thin and exhibited tidal
truncation, though the discs were able to expand to a somewhat larger
size in these cases. These features are indicated in the projection
plots for model~7 at $t= 328.5$ and model~8 at $t=228.0$ shown in
Fig.~(\ref{fig13}) and~(\ref{fig14}) respectively. Model~8 had $D/R=4$
and thus expanded to a larger radius until tidal effects became
significant.

\section{Discussion}

In this paper we have considered the nonlinear response of an
accretion disc to the tide of a binary companion when the disc
mid-plane is not necessarily coplanar with the plane of the binary
orbit. For our constant viscosity SPH models we found that tidal
truncation operates effectively when the disc and binary orbit are not
coplanar, being only marginally affected by the lack of
coplanarity. Our results indicate that modestly warped and thin discs
undergoing near rigid body precession may survive in close binary
systems, with the integrity of their local vertical structure
maintained. However, extremely thin discs may be severely disrupted by
differential precession and their survival depends only on their
hydrostatic state (i.e. the Mach number, ${\cal M}$, of the
unperturbed outer disc). For the binary mass ratios and separations
considered here, the crossover
between obtaining a warped, but coherent disc structure, and disc disruption,
appears to occur for values of the Mach number ${\cal M} \ge 30$. We also found
that the inclination evolved on a long timescale, likely to be the viscous
timescale, as indicated by the linear calculations of Papaloizou~\&
Terquem~(1995).  Below, we discuss some of the potential applications
of our models to a number of astrophysical phenomena of current
interest.

\subsection{Precessing jets in star forming regions}

Amongst the currently popular models formulated to explain the
generation of jets in young stellar objects, there are a class of
models that rely on the emission of a wind from the disc surface which
is then accelerated and collimated by the action of a magnetic field
(see for example K\"onigl~\& Ruden~1993 and references therein). It is
reasonable to assume that the wind will largely be emitted in a
direction normal to the disc surface, so that a precessing disc may
lead to the excitation of a precessing jet. The precession periods
obtained from our calculations with a mass ratio of 1 vary between 500
and 900 in our units. When scaled for a disc of radius 50 AU,
surrounding a star of $1 {\rm M}_{\odot}$, the unit of time is
$\Omega(R)^{-1} \simeq 56$~$yr$, leading to a precession period of
between $3.10^4$ and $5.10^4 yr$.

\noindent Bally~\& Devine~(1994) suggest that the jet which seems to be 
excited by the young stellar object HH34* in the L1641 molecular cloud
in Orion precesses with a period of approximately $10^4$~$yr$. This
period is consistent with the source being a binary with parameters
 similar to those we have used in our simulations with a separation on the
order of a few hundred astronomical units. We note that such binaries cannot
be resolved in Orion (at a distance of 470~pc) with current
ground based technology, but could be resolved with the Hubble Space Telescope.

\subsection{Spectral energy distributions of young stellar objects}

Some of the results presented in this paper demonstrate how a warped
disc can present a large  surface area for intercepting the primary
star's radiation. The effect that the consequent reprocessing of the stellar
radiation field can have on the emitted spectral energy distribution
has been investigated by Terquem~\& Bertout~(1993,~1996). They find
that the additional shrouding of the central star by a strongly warped
disc may account for the high spectral index of some T~Tauri stars or
even for the spectral energy distribution of some class~0/I
sources. Until now it was not clear to what extent a strongly warped
disc configuration could be created and maintained over long time
periods. However, the behaviour of Model~9 demonstrates that such a scenario
could  be physically realisable.

\noindent We note that the gradual warping of an accreation
disc will lead to it being progressively heated by the radiation from
the central source, so that the Mach number will then be a function of
time. Since the Mach number is the determining parameter in the global
evolution of the disc, we anticipate that the inclusion of this effect
may lead to changes in the evolution of the discs, particularly in the
case of model~9, where strong warping was observed to occur. Future
calculations will be presented to address this and related issues.

\subsection{Light curves of X--ray binaries}

There is evidence from the light curves of X--ray binaries such as
Hercules X-1 and SS433, that their associated accretion discs may be
in a state of precession in the tidal field
of the binary companion. We note that these systems are more complicated
than those modelled here because of
the process of mass transfer occuring
in them. In addition, numerical investigations on the accretion disc
coronae (Ko \& Kallman 1994) suggest disc mach numbers which our results
imply are marginal for survival under differential
precession in a tidal field. However, it is still of interest to compare
the precession
rates that are inferred observationally to those that we would predict
from our models. We assume that the discs around these objects are
tilted because of some internal process, for example the torque exerted by a
coronal wind (Horn~\& Meyer~1994), and that the secondary component 
induces a precession of the disc.

From equation~(\ref{Prc}), the ratio of the precession frequency to
the disc outer edge frequency is proportional to $(R/D)^3 \cos \left( \delta
\right) /q$, where $q=M_p/M_s.$ To obtain a general scaling we assume that the disc
radius is a fixed fraction of the mean Roche lobe radius. The ratio of the
Roche lobe radius $R_L$ to the separation is a function only of $q.$
We then find the approximate scaling for $q <1:$
\begin{displaymath}
\frac{P_{prec}}{P_{orb}} \propto
\sqrt{1+q} 
\left( 1+0.21 q^{1/3} \right)^{3/2} \frac{1}{\cos \delta},
\end{displaymath}
where $P_{orb}$ is the orbital period of the binary.
\noindent We calculate the factor of proportionality by using our 
numerical results. We find that $P_{prec}/P_{orb}$ varies between
10 and 20 as $q$ increases from 0 to 1 for small inclinations $\delta.$ 
Observations give a ratio close to 20.6 for Her~X-1 ($P_{orb}=1.7$ days and
$P_{prec}=35$ days, Petterson~1975, Gerend~\& Boynton~1976) and about 12.6 for
SS433 ($P_{orb}=13$ days and $P_{prec}=164$ days, Margon~1984 and references
therein). These values are consistent with the precession being induced by the
tidal field of the secondary.

\section*{Acknowledgments}
This work was supported by PPARC grant GR/H/09454, JDL is
supported by a PPARC studentship.
RPN and CT would like to thank the Astronomy
unit at QMW for hospitality.

\section*{References}

\noindent Artymowicz P., Lubow S.H., 1994, ApJ, 421, 651 \\

\noindent Bally J., Devine D., 1994, ApJ, 428, L65 \\

\noindent Bibo E.A., The P.S., Dawanas D.N., 1992, A\&A, 260, 293 \\

\noindent Corporon P., Lagrange A.M., Beust H., 1996, A\&A, 310, 228 \\

\noindent Edwards S., Cabrit S., Strom S.E., Heyer I., Strom K.M., Anderson E.,
1987, ApJ, 321, 473 \\

\noindent Gerend D., Boynton P., 1976, ApJ, 209, 562 \\

\noindent Gingold R.A., Monaghan J.J., 1977, MNRAS, 181,375 \\

\noindent Goldreich P., Tremaine S., 1981, ApJ, 243, 1062 \\

\noindent Hernquist L., Katz N., 1989, ApJ Sup. S., 70, 419 \\

\noindent Horn S., Meyer F., 1994, in Duschl W.J., Frank J., Meyer F.,
Meyer-Hofmeister E., Tscharnuter W.M., eds, Theory of Accretion
Disks-2, NATO ASI Series (Kluwer: Dordrecht), p. 163 \\

\noindent Ko Y., Kallman T.R., 1994, ApJ, 431, 273 \\

\noindent K\"onigl A., Ruden S.P., 1993, in Levy E.H., Lunine J., eds,
Protostars and Planets III (Univ. Arizona Press, Tucson), p. 641 \\

\noindent Leinert C., Zinnecker H., Weitzel N., Christou J., Ridgway S.T.,
Jameson R.F., Haas M., Lenzen R., 1993, A\&A, 278, 129 \\

\noindent Lin D.N.C., Papaloizou J.C.B., 1993, in Levy E.H., Lunine J., eds,
Protostars and Planets III (Univ. Arizona Press, Tucson) p.749 \\

\noindent Lucy L.B., 1977, AJ, 83, 1013 \\

\noindent Lynden-Bell D., Ostriker J.P., 1967, MNRAS, 136, 293 \\

\noindent Lynden-Bell D., Pringle J.E., 1974, MNRAS, 168, 603 \\

\noindent Margon B., 1984, ARA\&A, 22, 507 \\

\noindent Mathieu R.D., 1994, ARA\&A, 32, 465 \\

\noindent Monaghan J.J., Gingold R.A., 1983, J. Comp. Phys., 52, 374 \\

\noindent Monaghan J.J., Lattanzio J.C., 1985, A\&A, 149, 135 \\

\noindent Nelson R.P., Papaloizou J.C.B., 1993, MNRAS, 265, 905 \\

\noindent Nelson R.P., Papaloizou J.C.B., 1994, MNRAS, 270, 1 \\

\noindent Nelson R.P., 1994, Ph.D Thesis, University of London \\
 
\noindent Papaloizou J.C.B., Lin D.N.C., 1995, ApJ, 438, 841 \\

\noindent Papaloizou J.C.B, Pringle J.E., 1983, MNRAS, 202, 1181 \\

\noindent Papaloizou J.C.B., Terquem C., 1995, MNRAS, 274, 987 \\

\noindent Petterson J.A., 1975, ApJ, 201, L61 \\

\noindent Savonije G.J., Papaloizou J.C.B., Lin D.N.C., 1994, MNRAS, 268, 13 \\
 
\noindent Shakura N.I., Sunyaev R.A., 1973, A\&A, 24, 337 \\

\noindent Spruit H.C., 1987, A\&A, 184, 173 \\

\noindent Terquem C., Bertout C., 1993, A\&A, 274, 291 \\

\noindent Terquem C., Bertout C., 1996, MNRAS, 279, 415 \\

\newpage 

\begin{figure}
\caption{Projection plots in orbital plane coordinates for model~1 
at time $t \simeq 0$. The projections are in the $(x,y_o)$ plane (top
left), $(x,z_o)$ plane (top right), $(y_o,z_o)$ plane (bottom left)
and $(x,z_o)$ plane (bottom right).}
\label{fig1}
\end{figure}

\begin{figure}
\caption[]{The precession angle $\beta_p$ (dashed line) and inclination angle 
$\iota$ (solid line) for model~1.}
\label{fig2}
\end{figure}

\begin{figure}
\caption{Projection plots in orbital plane coordinates for model~1
at time $t=156.6$.}
\label{fig3}
\end{figure}

\begin{figure}
\caption{Projection plot in the initial disc plane $(x,y)$
for model~1 at time $t=0$ and $t=222.8$. The disc at $t=222.8$ is seen
almost edge-on.}
\label{fig4}
\end{figure}

\begin{figure}
\caption{Projection plots in orbital plane coordinates for model~1
at time $t=297.8$.}
\label{fig5}
\end{figure}

\begin{figure}
\caption{Projection plots in orbital plane coordinates for model~3
at time $t=330.0$.}
\label{fig6}
\end{figure}

\begin{figure}
\caption{Projection plots in orbital plane coordinates for model~9 
at time $t=397.9$.}
\label{fig7}
\end{figure} 

\begin{figure}
\caption{The precession angle $\beta_p$ (solid and long-dashed lines) 
and inclination angle $\iota$ (short-dashed and dotted lines) for
model~9. The solid and short-dashed lines correspond to the outer disc
section with $r>0.5$, and the long-dashed and dotted lines correspond
to the inner disc section with $r<0.3$.}
\label{fig8}
\end{figure}

\begin{figure}
\caption{The precession angle $\beta_p$ (solid and long-dashed lines) 
and inclination angle $\iota$ (short-dashed and dotted lines) for
model~2. The solid and short-dashed lines correspond to the outer disc
section with $r>0.5$, and the long-dashed and dotted lines correspond
to the inner disc section with $r<0.3$.}
\label{fig8bis}
\end{figure}

\begin{figure}
\caption{Projection plots in orbital plane coordinates for model~4 
at time $t=105.1$.}
\label{fig9}
\end{figure} 

\begin{figure}
\caption{The precession angle $\beta_p$ (dashed line) and inclination 
angle $\iota$ (solid line) for model~4.}
\label{fig10}
\end{figure} 

\begin{figure}
\caption{Projection plots in orbital plane coordinates for model~11 
at time $t=147.3$.}
\label{fig11}
\end{figure} 

\begin{figure}
\caption{Projection plots in orbital plane coordinates for model~6 
at time $t=84.0$.}
\label{fig12}
\end{figure} 

\begin{figure}
\caption{Projection plots in orbital plane coordinates for model~7 
at time $t=328.5$.}
\label{fig13}
\end{figure} 

\begin{figure}
\caption{Projection plots in orbital plane coordinates for model~8 
at time $t=228.0$.}
\label{fig14}
\end{figure} 

\end{document}